\documentstyle[aps,preprint]{revtex}
\begin{document}
\title{Can Extremal Black Holes Have Non-Zero Entropy ?}
\author{Saurya Das, Arundhati Dasgupta and P. Ramadevi
\footnote{E-Mail:~saurya,dasgupta,rama@imsc.ernet.in}}
\address{The Institute of Mathematical Sciences, \\ CIT Campus,
Madras - 600 113,  India.}
\maketitle
\begin{abstract}
We give several pieces of evidence 
to show that extremal black holes cannot be 
obtained as limits of non-extremal black holes. 
We review arguments in the literature showing that the entropy of 
extremal black holes is zero, while that of near-extremal ones obey 
the Bekenstein-Hawking formula. 
However, from the counting of degeneracy of quantum
(BPS) states of string theory the entropy of
extremal stringy black holes obeys the area law.
An attempt is made to reconcile these arguments. 
\end{abstract}
%\newpage

\section{Introduction}
Black hole thermodynamics has been an active area of research since
it was shown that the area of the event horizon of a black hole can
be identified (modulo constants) with the `entropy' of the black hole.
The precise relation is \cite{bek,haw}:
\begin{equation}
S_{BH}~=~ \frac{A}{4G}~,
\label{bh}
\end{equation}
where $S_{BH}$ stands for the Bekenstein-Hawking entropy, $A$ is
the horizon area and $G$ is Newton's constant. 
Ever since, it has been one of the outstanding 
problems to attribute $S_{BH}$
to the degeneracy of quantum states of the black hole. 
There have been
several attempts in this direction and recently some
significant work has been done in this area
\cite{sen,vafa,hor}. 

It may be noted that, if one considers black holes with either charges or
angular momentum or both, then
the relation (\ref{bh}) is valid only for non-extremal black holes. 
However, extremal black holes occupy a special place in
black hole physics. They have $T_H=0$, ensuring 
stability against Hawking radiation. 
Several derivations have been given which demonstrate
semi-classically that the entropy
of extremal Reissner-Nordstr\"om black holes is actually zero \cite
{ross,haw2,teitel}. This
means that there is a unique quantum microstate corresponding to these
black holes. 
Thus, extremal black holes cannot be
regarded as limits of non-extremal black holes 
because of a discontinuity at that limit. 
The underlying reason is that non-extremal and
extremal black holes are topologically different objects and one
cannot change the topology continuously. 

On the other hand, string theory predicts that the entropy of
extremal black holes follows the relation (\ref{bh}), by suitably
identifying these black holes with elementary and/or solitonic string
states at weak string coupling \cite{sen,vafa,hor}. Counting the
degeneracy of these states and taking its logarithm reproduces
$A/4G$ exactly \footnote {In certain cases
$A$ might correspond to the area of the `stretched horizon'
\cite{sen,peet}.}. Thus we are faced
with the question : what is the entropy of an extremal black
hole? While the proofs given in \cite{ross,teitel} seem quite robust,
they are nevertheless semi-classical. Can string theoretic
corrections near the horizon modify their result to produce $S= A/4G$?
We address this question in this paper
and pose a possible way of resolving this problem. 

We first show in section \ref{ent} as to why the
entropy of extremal Reissner-Nordstr\"om 
black holes cannot be given by (\ref{bh}). Next, we summarize the
arguments leading to the result that extremal Reissner-Nordstr\"om
black holes have zero entropy \cite{ross,teitel}. This implies that
there is a discontinuous change in the BH entropy at the extremal limit. 
We investigate this further in section \ref{phy} to check  
whether a non-extremal black hole can
be transformed into an extremal one by any physical process.  
In particular, we consider Hawking radiation, 
Penrose process and super-radiance, by which black holes can lose
both mass and charge. 
In Penrose process and
superradiance  
particles scattering from non-extremal black holes carry away charge and
mass from the latter. 
We show that these processes cannot transform the
black hole into an extremal one. 
Next, considering a non-extremal black hole as a perfect
blackbody, radiating with a characteristic Hawking temperature 
$T_H$, we show that the time taken for it to radiate and reach
extremality is infinitely large. In other words, the black hole
never becomes an extremal one by Hawking radiation. 
These
evidences seem to confirm that extremal black holes are
physically quite different from non-extremal ones. 
In section \ref{string}, we examine the extremal stringy black
holes considered recently in the context of BH entropy. Following
the reasoning of section \ref{ent} the entropy of these 
black holes must be zero. However, we  propose a modification
of the metric and show that it corresponds to the extremal limit
of the stringy black hole for which the area law is valid.

%\newpage
\section{Entropy of Extremal Black
Holes}
\label{ent}

Let us consider the four dimensional Reissner-Nordstr\"om (RN) metric;
\begin{equation}
ds^2~=~-\left(1 - \frac{2GM}{r} + \frac{GQ^2}{r^2} \right) dt^2
+ \left(1 - \frac{2GM}{r} + \frac{GQ^2}{r^2} \right)^{-1} dr^2
+ r^2 d\Omega^2~.
\end{equation}
The following relation holds when the black hole 
undergoes an infinitesimal changes in mass,
charge and horizon area :
\begin{equation}
dM~=~\frac{k}{8\pi G} dA + \phi dQ~,
\label{bhtd}
\end{equation}
where $M$ and $Q$ are the mass and the charge of the black hole
respectively. The horizons are at $r_\pm = GM \pm G \sqrt{M^2 -
Q^2/G}$. The surface gravity $\kappa$ is given by
$(r_+-r_-)/2r_+^2$ and $\phi = Q/r_+$ is the electrostatic scalar
potential at the outer (event) horizon. The Hawking temperature of
the black hole is given by $T_H=\kappa/2\pi$ and $M$ is just the
energy $E$ of the black hole. Comparing with the first law of
thermodynamics,
\begin{equation}
dE~=~ TdS - PdV~,
\label{td}
\end{equation}
and replacing $PdV$ by $-\phi dQ$, we see that the entropy $S$
must be identified with $A/4G$ \footnote{upto an additive
constant , which we set to zero by demanding that $S_{BH}
\rightarrow 0$ as $M \rightarrow 0$}, thus giving Eq.(\ref{bh})
\cite{haw}.

Now, for extremal RN black holes, $T_H=0$, since $r_+=r_-$. Thus,
we can no longer compare equations (\ref{bhtd}) and (\ref{td}) to
obtain (\ref{bh}). Moreover, using the relation for temperature
\begin{equation}
T_H^{-1}~=~\left(\frac{\partial S_{BH}}{\partial M}\right)_Q~,
\end{equation}
we see that the right hand side diverges at $T_H=0$, indicating
that the entropy as a function of $M$ has a singularity at the
extremal limit $M=Q/\sqrt{G}$. The above arguments hold good for
$D$-dimensional charged black holes as well, for which 
$T_H = (D-3) (r_+^{D-3}-r_-^{D-3})/4\pi r_+^{D-2}$. 

Next, we analyze an alternative derivation of Eq.(\ref{bh}) 
given in \cite{bek}, for the case of
extremal RN black hole. The entropy is assumed to be an arbitrary
function of area:
\begin{equation}
S_{BH}~=~f(A)~,
\end{equation}
from which, one can write
\begin{equation}
\Delta S_{BH}~=~\frac {d~f}{d A} \Delta A~,
\label{sbh}
\end{equation}
where $\Delta S_{BH}$ and $\Delta A$ correspond to the
change in the entropy and area respectively of the black
hole when a particle falls into it. The quantity
$\Delta S$ can be found from the point of view of 
information theory. Before it enters the event horizon, 
it is certain that the particle exists. Once it enters
the horizon, we have no information whatsoever about
the particle, and it is justified to assume that
it is equally probable for it to exist
or not. Thus the minimum entropy
change (ignoring possible internal structures of the
particle) is given by
\begin{equation}
(\Delta S)_{min}~=~\sum_n~p_n \ln p_n~=~ \ln2~,
\end{equation}
where summation over $n$ corresponds to all possible
states of the particle. 
Now, for an infalling particle with mass $\mu$ and
its center of mass at $r_+ + \delta$, the
proper radius 
$b$ is $\int_{r_+}^{r_+ + \delta} \sqrt {g_{rr}} dr$. Then the
minimum change of black hole area
accompanied by the absorption of the 
particle will be \cite {bek}
\begin{eqnarray}
(\Delta A)_{min}~=~ 2 \mu b~, ~~~~ {\mbox with }\nonumber\\
b~=~2 \delta^{1/2} \frac {r_+}{\sqrt {r_+-r_-}}~.
\end{eqnarray}
where $b$ is obtained using 
$g_{rr}~=~ (r-r_-)(r-r_+)/r^2$, 
and non-extremality condition $(r_+ - r_-\gg \delta)$. 
However, in the extremal limit $(r_+\rightarrow r_-)$, we get:
\begin{equation}
b~=~\delta~+~r_+\ln(r-r_+)|_{r_+}^{r_+ + \delta}~,
\label{bmin}
\end{equation}
which diverges for any $\delta >0$. This means that
for any finite $\delta$, however small, the corresponding proper
radius of the infalling particle is infinite. Thus, the above
equation makes sense only  for $\delta~=~0$. Thus, we take
$b~=~0$ corresponding to a point particle
resulting in $(\Delta A)_{min}~=~0$.
Thus to satisfy equation (\ref{sbh}) we require,
\begin{equation}
\left (\frac {\partial f}{\partial A} \right )_{r_+=r_-} \longrightarrow \infty~,
\end{equation}
which once again shows that the entropy is not continuous at the
extremal limit.

The discontinuous nature of entropy under the transition
from non-extremal to extremal black hole asserts that the
entropy of extremal black holes cannot
be determined as a limit of the non-extremal one.
Independent derivations of the entropy for extremal and non-extremal black
holes  have been given \cite {teitel} which are in conformity with
the above result. 
It has been shown that the topology
of the black hole near the horizon plays a crucial
role in determining the entropy. We now briefly
review these arguments.

The Euclideanised metric in $d$ dimensions near the
horizon is
\begin{equation}
ds^2~=~ N^2 d\tau^2 + N^{-2} dr^2 + r^2~d \Omega_{d-2}^2~.
\end{equation}
For the above metric, the proper angle $\Theta $
in the $r- \tau$ plane near the horizon is defined as
\begin{equation}
\Theta~\equiv~\frac {\mbox{proper length}}{\mbox{proper radius}}~=~
\frac{\int_{t_1}^{t_2} {\sqrt {g_{\tau \tau}}~d\tau}}
{\int_{r_+}^{r} \sqrt{g_{rr}}~dr}~=~ (N N')|_{r_+} (t_2-t_1),
\label{proper}
\end{equation}
where the prime denotes differentiation with respect to $r$.
It can be shown that
$N$ satisfies the following relation:
\begin{equation}
(t_2-t_1) N^2~=~2\Theta~(r-r_+) + O [(r-r_+)^2]~.
\end{equation}
Also, the two dimensional 
metric near the horizon can be written in the form 
\begin{equation}
ds^2~=~d\rho^2 + \rho^2 d\Theta^2~,
\end{equation}
where $\rho \equiv  \sqrt{2 (r-r_+)/NN'}.$ 
To avoid a conical singularity at the horizon, the period of 
$\Theta$ is identified with $2\pi$,  which corresponds to the
topology of a {\it disc} with zero deficit angle 
in the $r-\tau$ plane. This can always be done for non-extremal
black holes, as $(NN')|_{r_+}$ in Eq.(\ref{proper}) is non-zero.
However, for
extremal black holes, the proper radius diverges (see Eq.(\ref
{bmin})), and hence the proper angle tends to zero. 
Thus the conical deficit
angle becomes $2\pi$ and the topology is that of an {\it annulus}
\cite{teitel}. The topology of the transverse section in
either case is $S^{d-2}$.

Now, we need to see how this topology reflects on the
entropy calculation. Treating the black holes as
microcanonical ensemble, the action $I$ in the
Hamiltonian formulation of gravity is  proportional
to entropy. The dimensional continuation of  Gauss-Bonnet
theorem to  $d$ dimensions \cite {teitel,teitel1} determines 
\begin{equation}
I~\propto~ \chi ~ A_{d-2}
\label{act}
\end{equation}
where $\chi$ is the euler characteristic of the Euclideanised
$r- \tau$ plane and $A_{d-2}$ is the area of the transverse
$S^{d-2}$. 
The exact expression for the black hole entropy is given by
\begin{equation}
S~=~\frac{ \chi A}{4G}~.
\end{equation}
For non-extremal black holes $\chi =1$ (disc), leading to the area law
(\ref{bh}), while for extremal black holes $\chi=0$ (annulus), implying a
vanishing entropy. 

It has also been argued by Hawking et al \cite {ross,haw2},
that the Euclidean action for extremal black
holes is proportional to the inverse Hawking
temperature ($\beta$) in a canonical ensemble leading to 
the vanishing entropy. This follows from the relations
$S~=~ -\left( \beta \frac{\partial}{\partial \beta} - 1\right) \ln
Z~,$ and $ Z~=~e^{-I}~.$ \footnote {Extremal black hole entropy has been
explored by alternative methods as well in \cite{gh}.} 

Thus, it is clear that extremal black holes cannot be thought 
of as limits of non-extremal black holes at least as far as the
expression for their entropies are concerned. In the next
section, we investigate some physical processes
which further support this conclusion. 

\section{Physical Processes}
\label{phy}

For the charged non-extremal black holes, we know that
Hawking radiation is dominant in the energy regime
$\omega >  e \phi$ where $e$ is the charge of the
emitted particles. On the other hand, the Penrose
process (and its quantum analog - superradiance)
is significant when $\omega < e \phi$. We study
both the processes, thus spanning 
all the energy regimes, to confirm that non-extremal
black holes cannot transform into extremal ones.

\subsection{Superradiance and extremality}
\label{super}
RN black holes can lose mass and charge by  processes like
Penrose process and superradiance, which are dominant for low 
energies of the infalling particles. We examine here, whether 
a non-extremal black hole can reach extremality through
these processes.

The energy of a particle in a $4D$ RN background is given by
\cite{wald}
\begin{equation}
E=m \sqrt{\left(1-{r_+\over r}\right)\left(1-{r_-\over r}\right)} + {eQ\over {r}}~.
\end{equation}
Where $m$ and $e$ are the mass and charge of the infalling particle.
If this particle has a charge opposite to that of the black hole,
then sufficiently close to the horizon, the first term tends
to zero, making the energy negative. Hence in this regime, 
\begin{equation}
|E|<{|eQ|\over {r_+}}~.
\label{eq:cond}
\end{equation}
If two oppositely charged bound particles 
with total energy $E_{0}$ fall near the black hole and separate
there, one of the charges can have negative energy by the
above argument.
The particle with negative energy will fall into the black hole
and the other particle escapes. 
By conservation of energy,
\begin{eqnarray}
E'_{2}&=&E_{0}+|E_{1}|\\
M'&=&M -|E_{1}|
\label{eq:mas}
\end{eqnarray}
$E'_{2}$ is the energy of the particle which escapes, $E_{1}$
is the energy of the particle which falls into the black hole and 
$M'$ is the final mass of the black hole. Thus there is a decrease
in mass of the black hole, while the escaping particle carries
back more energy. Also as an oppositely charged
particle is absorbed by the black hole, it's effective charge
decreases to become $ Q'=Q-e$
Since for a non-extremal black hole (${{\sqrt G} Q/{r_{+}}}<1$). We find from
equation(\ref{eq:cond}), that 
\begin{equation}
{e\over {\sqrt G}}>|E_{1}|~. 
\label{eq:co}
\end{equation}
In other words 
the decrease of mass of the black hole will be less than the decrease
of charge due to this process. Hence, the condition of non-extremality
$M>{Q/{\sqrt G}}$ will be maintained as the rate of charge loss will exceed
the rate of mass loss.

The quantum analog for this phenomenon is Superradiance. Fields
with low energy are shown to be scattered away from the black
hole such that the reflection coefficient is greater than one.
The charged scalar field equation can be solved in the RN back ground
and the following relation for the reflection coefficient $|R|^2$ and
transmission coefficient or the absorption coefficient $|T|^2$,
can be obtained as \cite{gib},
\begin{equation}
1 - |R|^2= {1\over {k}}(\omega -{eQ\over {r_{+}}})~|T|^2~.
\end{equation}
For $\omega< eQ/r_{+}$, the reflection coefficient, $|R|^2$ is
greater than 1, or the scalar wave takes away energy from the
black hole. The condition for superradiance is thus
\begin{equation}
m<\omega< \frac{eQ}{r_+}~.
\label{eq:co1}
\end{equation}
The rate of charge loss and mass loss for the black hole is 
\begin{eqnarray}
{dQ\over {dt}}&= &-e \int_{m}^{eQ\over r_{+}}{ |R|^2}d\omega\\
{dM\over dt}&=& -\int_{m}^{eQ\over r_{+}}{|R|^2}\omega d\omega~.
\end{eqnarray}
We find that the for the initial value of the integrands, $(e/{\sqrt
G})|R(m)|^2 > m|R(m)|^2$. Thus as equation (\ref{eq:co1}) holds for each value of $\omega$,which is bounded from above, 
\begin{equation}
|{dQ\over dt}|>{\sqrt G} |{dM\over dt}|~.
\end{equation}
Hence from the quantum process also it is clear that the $M>{Q/
{\sqrt G}}$
condition will be maintained. 

The above result is easily extendible to higher dimensional
charged dilatonic black holes. The equation for
classical energy of a charged particle in a generic charged black hole
back ground is 
 \footnote{for the $D$-dimensional charged metric, see \cite{maeda}.}
\begin{equation}
E=m \sqrt{\left(1-({r_+\over
r})^{D-3}\right)\left(1-({r_-\over
r})^{D-3}\right)^{1-{2a^2/(D-3 + a^2)}}} + {eQ\over
{r^{D-3}}}~.
\end{equation}
Here $D$ is the dimension of space and $a$ stands for a
parameter which interpolates between the general relativistic
solution $a=0$ and the dilatonic stringy black hole $a=1$.
As $r_+^{D-3}= GM + G \sqrt{M^2 - Q^2/G}$, 
non-extremality will
imply ${\sqrt G}Q<r_{+}^{D-3}$ and
equation (\ref{eq:co}), hold for these. 
The equation for the reflection and transmission coefficients
for these black holes has been calculated in {\cite{kiy}}. The
condition for superradiance equation
({\ref{eq:co1}}), is the same for these black holes.
 
Apart from the induced process stated above, the black hole loses
charge spontaneously by vacuum polarization as shown in {\cite{gib}}.
For this the rate of charge loss will also be very high, and
the black hole will tend to discharge itself very fast. The
$M=Q/{\sqrt G}$ condition, once again, will not be obtained.
Thus in the processes considered so far, the extremality condition
cannot be attained from a non-extremal state.

\subsection{Hawking Radiation and Extremality}
\label{rad}
In this section, we consider mass and charge loss of black holes
by Hawking radiation. 
 When the RN black hole radiates, the spectrum of particles 
 is given by the Planck distribution \cite{haw}:
\begin{equation}
dE_\omega~=~\frac{(\omega-e\phi)^3~d\omega}{e^{(\omega-e\phi)/T_H} -1 }~,
\end{equation}
where $dE_\omega$ is the radiation energy in the spectral range
$\omega$ to $\omega + d \omega$.
Integrating over $\omega$ from
$e\phi$ to $\infty$, one
obtains the rate at which the black hole loses energy, i.e. mass
\cite{wald}
\begin{equation}
\frac{dM}{dt}~=~-\sigma T_H^4~A~,
\end{equation}
$A=4\pi r_+^2$ being the area of the event horizon and $\sigma$ 
the Stefan-Boltzmann constant. Thus, for RN black holes with
$T_H=(r_+-r_-)/4\pi r_+^2$, it is given by,
\begin{equation}
\frac{dM}{dt}~=~-\frac{\sigma}{(4\pi)^3}\frac{(r_+-r_-)^4}{r_+^6}~.
\label{sig}
\end{equation}
We integrate (\ref{sig}) to get, 
\begin{equation}
\int_0^{t_0}~dt~=~ - \frac{(4\pi)^3}{\sigma}\int_{M_0,Q_0}^{M',Q'}~\frac{r_+^6~
dM}{(r_+-r_-)^4}~.
\label{int}
\end{equation}
Here $t_0$ is the time taken for the black hole to reach a final
state with mass and charge $M'$ and $Q'$ respectively, from their
initial values $M_0$ and $Q_0$. We are interested 
in calculating $t_0$ to reach  a final extremal state (i.e. $M'=Q'/{\sqrt
G}$). 
from a non-extremal initial state.  
For simplicity, let us first
assume that the radiated particles are electrically neutral, i.e.
$e=0$ and hence $Q$ is a constant. Then, the time taken for the black
hole to become extremal is:
\begin{eqnarray}
\int_0^{t_0}~dt~&=&~-\frac{(4\pi)^3}{\sigma}~\lim_{M' \rightarrow Q_0/{\sqrt G}}~
\int_{M_0}^{M'}~\frac{r_+^6~
dM}{(r_+-r_-)^4}~,  \nonumber \\
&=&~\frac{(4\pi)^3}{\sigma}\lim_{M' \rightarrow Q_0/{\sqrt G}}~
\left[\frac{105}{12}Q_0^3 {\sqrt G}\ln\left(\frac{M+Q_0/{\sqrt G}}{M-Q_0/{\sqrt G}}
\right) \right. \\ 
&& \nonumber \\
&+&
~\left( \frac{328Q_0^4  -128 M^4 G^2-
128 Q_0^2 M^2 G}{12(M^2 - Q_0^2/G)^{1/2}}
\right) 
+ \left.\left .  \left( \frac{ 198 Q_0^4 M-128 M^5 G^2
- 64 Q_0^2 M^3 G}{12 (M^2 - Q_0^2/G)}  
\right)\right] \right|_{M=M_0}^{M=M'}~.  \nonumber 
\end{eqnarray}
Clearly, $t_0$ diverges. That is, the RN black hole which emits
neutral particles, takes an
infinite amount of time to reach extremality. Generalizing the
proof for Hawking particles carrying charges is not difficult. 
Then $Q$ is not a constant in Eq.(\ref{int}). 
However, as before, the integrand on the right hand side diverges
as $r_+ \rightarrow r_-$ and thus $t_0 \rightarrow \infty$. 
Identical conclusions follow for general relativistic 
charged black holes in $D$-dimensions, for which, the rate of
mass loss is given by 
$$\frac{dM}{dt}~=~-\sigma_D~A_{D-2}\left(\frac{D-3}{4\pi}\right)^D~
\frac{[r_+^{D-3}-r_-^{D-3}]^D}{r_+^{(D-2)(D-1)}}~,$$
$\sigma_D$ being the $D$-dimensional Stefan-Boltzmann constant and
$A_{D-2}$ the area of unit $S^{D-2}$. Here
too the integrand diverges in the extremal limit. In general, $t_0
\rightarrow \infty$ whenever $T_H=0$ for the extremal black hole.
We shall see later, that 
this includes a certain class of stringy black holes.

Thus, we conclude that a extremal black hole state with $T_H=0$
cannot be reached in a finite time by Hawking radiation 
from a non-extremal black hole. This is in conformity 
with the third law of black hole thermodynamics, which 
asserts that the same cannot be reached in
a finite sequence of operations \cite{carter}. 
These also provide pieces of evidence that the area law for
the entropy of non-extremal black holes cannot be extended to the
case of extremal black holes. In the next section, we will study
the entropy of certain extremal stringy black holes which supposedly
obey the area law. 

\section{Extremal Black Holes in String Theory}
\label{string}
The  degeneracy counting of string states saturating
the BPS bound has been claimed to give
the entropy of the extremal stringy black
holes  with the same mass and charge. The surprising fact is that the
entropy obtained by degeneracy counting matches
the area law, which is applicable only to non-extremal black holes.
We will try to reconcile this apparent
discrepancy in this section.

There are two types of extremal stringy black holes
which saturate the BPS bound :
\begin{enumerate}
\item  The horizon merges with the curvature 
singularity. These black holes have zero horizon area
and the dilaton field becomes singular at the horizon.
\item The two event horizons coincide as in General relativity. 
The area for these black holes is non-zero and the dilaton 
is regular at the horizon.
\end{enumerate}
A few examples of these extremal stringy black holes and
their properties are tabulated below.

\vspace{.5cm}
\begin{center}

\begin{tabular}{||c|c|c|c|c|c|c|c||} \hline
Type&Example&\multicolumn{2}{c|}{$T_H$}  & \multicolumn{2}{c|}{Macro Entropy}& 
\multicolumn{2}{c|}{Micro Entropy} \\ \cline{3-8}
{} & {} & ~~NE~~ &E& ~~NE~~ & ~E & ~~NE~~ & E \\  \hline
1&Het. on $T^6$&$\neq 0$&$1/4\pi m_0$&$A/4$& $ 0 $ & -& $ A_{st}/4$ \\ 
\hline
2&II B on $K^3 \times S^1$&$ \neq 0$&$ 0$ & $A/4$ &$0$ &-& $A/4$  \\ \hline
\end{tabular} \\

\end{center}
\vspace{.5cm}
where NE = nonextremal, E = extremal  and $A_{st}$ is
the area of the stretched horizon. The examples 
referred to here are taken from  Refs. \cite {sen,vafa}. 

The first type of extremal black holes \cite {sen}, obtained
by compactifying heterotic on $T^6$ has 
$T_H \neq 0$. Hence in accordance with the 
third law of black hole thermodynamics, the extremal
state can be reached in a finite sequence of steps. 
In particular, one can see from 
Section \ref {rad}, that the time $t_0$ to reach extremality by
Hawking radiation is finite 
\footnote {The grey-body factor for the metric (\ref{str}),
for low energy quanta, seems to be zero because of vanishing horizon area.
However, as argued in \cite {sen},
this metric suffers
large stringy corrections resulting in the 
metric (\ref{sth}) which has a non-zero
(stretched) horizon area.}.
Similarly, the discussions 
in Section \ref {ent} based on \cite {bek,haw} is valid  for this
example since proper radius is finite. 
It follows that these extremal black
holes can be regarded as limits of non extremal ones
and their entropy obeys the area law. 
In order to determine this entropy, we look at the
extremal stringy black hole solution of the low
energy effective action of heterotic string theory compactified on
$T^6$. The Euclideanised metric near the horizon is \cite{sen}:
\begin{equation}
d {\bar s}^2 ~=~ \frac{1}{4}{\bar r}^2 d{\bar \tau}^2 + d{\bar r}^2 + \frac{1}{4} {\bar
r}^2 \left( d\theta^2 + \sin^2\theta d\phi^2\right)~,
\label{str}
\end{equation}
and the solution for the dilaton field is 
$e^{\bar \phi}~=~{\bar r}^2/4~.$ Here $\bar r^2= 4 g r$
and $\bar \tau=\tau /m_0$ where $g$ is the string 
coupling and the parameter $m_0$ is related to the
mass of the black hole.
Note that the topology near the horizon ($\bar r=0$)
is {\it disc} $\times S^2$, which is that of a generic
non-extremal black hole. Although $S=A/4G$, the entropy vanishes
as the horizon area is zero. On the other hand, the 
degeneracy counting of the elementary BPS
string states gives a non-zero result. It has been proposed that
stringy corrections near the horizon modifies the metric and the
dilaton such that the results agree. A possible modification of
the metric and the dilaton is
\begin{equation}
d {\bar s}^2 ~=~ \frac{1}{4}{\bar r}^2 d{\bar \tau}^2 + 
d{\bar r}^2 + \frac{1}{4} f_1(\bar r) \left( d\theta^2 + \sin^2\theta d\phi^2\right)~~;~~
e^\phi~=~f_2(\bar r)~,
\label{sth}
\end{equation}
where $f_1(\bar r)$ and $f_2(\bar r)$ are two smoothing
functions which are positive constants at $\bar r=0$ and 
equal $\bar r^2 /4$ for large $\bar r$. Now, the horizon 
area can be shown to be  $\sqrt {m_0 f_1(0)/g}$, which is
finite and proportional to the logarithm of the number 
of corresponding elementary string states satisfying the 
BPS bound. Here $f_1(0)$ is obtained by a fit with the
degeneracy.

The second type of the stringy extremal black holes 
obtained from type IIB string theory compactified on $K^3 \times
S^1$, on the contrary, has $T_H=0$. The horizon area $A \neq 0$ 
if both NS-NS  charge $Q_H$ and R-R charge $Q_F$,
associated with the $\tilde H$ and $F$ field strengths 
respectively, are non-zero \cite {vafa}. 
The extremal black hole solution from low energy effective theory
is the five-dimensional RN metric \cite {maeda}: 
\begin{equation}
ds^2~=~-\left(1-\left(\frac{r_0}{r}\right)^2\right)^2~dt^2 +
\left(1-\left(\frac{r_0}{r}\right)^2\right)^{-2}~dr^2 + r^2
d\Omega_3^2~.
\label{5d}
\end{equation}
where the horizon radius in terms of the charges $Q_H$ and $Q_F$ is
$r_0=\left(8 Q_H Q_F^2/ \pi^2
\right)^{1/6}$. As discussed in Section \ref {ent},
the Euclidean topology is {\it annulus}
$\times S^3$ and hence its entropy is zero. 
The degeneracy counting has been done by identifying
a collection of BPS saturated D-brane states 
in weak coupling regime with the extremal 
charged black holes at strong coupling.
The logarithm of this degeneracy exactly matches the area 
law for the metric (\ref{5d}). The method was applied to certain four
dimensional stringy black holes as well, with identical conclusions
\cite{mal}.

The previous discussions suggest that there could be string theoretic
or other quantum gravity corrections which would prevent 
the metric near the horizon from being {\it exactly}
extremal, such that the area law continues to be valid. 
It has been argued that Planck scale effects become important 
near the horizon \cite {suss,thf}.
Stringy modifications were also anticipated in \cite{hor} 
on the basis of stability requirements.
In view of the above, the modified metric with the correct
topology should be of the form,
\begin{equation}
ds^2~=~-f(r)~\left(1-\left(\frac{r_0}{r}\right)^2\right)~dt^2 +
f(r)^{-1}\left(1-\left(\frac{r_0}{r}\right)^2\right)^{-1}~dr^2 + r^2
d\Omega_3^2~,
\label{5d'}
\end{equation}
where $f(r)$ is a positive definite
and bounded function of $r$ in the range $r_0 \le r<\infty$,
such that $f(r_0) \neq 0$, although it can be arbitrarily
small. The corresponding Hawking temperature gets modified
from zero to $T_H = f(r_0) /2\pi r_0$. 
 
However though the exact nature of the stringy corrections are
not ascertained, the metric (\ref{5d'}) can be understood from
an alternative approach. 
In general the metric solution of type IIB action 
compactified on  five
dimensional manifold is of the form \cite{udu}:
\begin{equation}
ds^2~=~-f^{-2/3} \left(1 - \frac{a^2}{\bar r^2}\right) + f^{1/3}
\left[ \left( 1 - \frac{a^2}{\bar r^2} \right)^{-1} d\bar r^2 + r^2
d\Omega_3^2 \right]~,
\end{equation}
where,
\begin{equation}
f~=~\left( 1 + \frac{a^2 \sinh^2\alpha}{\bar r^2} \right)
~\left( 1 + \frac{a^2 \sinh^2\gamma}{\bar r^2} \right)
~\left( 1 + \frac{a^2 \sinh^2\sigma}{\bar r^2} \right)
\end{equation}
Here, $\alpha$, $\gamma$ and $\sigma$ are three boost parameters. These 
parameters along with $a$, radius $R$ of $S^1$ and volume $V$ of the four 
dimensional compact manifold determine the number of 
one branes, five branes, the corresponding anti-branes  
and the momentum in the $S^1$ direction. The five
dimensional Reissner-Nordst\"orm metric is obtained when all
these boost parameters are equal, i.e.
$ \alpha = \gamma = \sigma~,$
and simultaneously doing  the coordinate transformation 
$ r^2~=~\bar r^2 + 
a^2 \sinh^2 \alpha~$:
\begin{eqnarray}
ds^2~&=&~-\left[ 1 - \frac{(a \sinh^2 \alpha)^2}{r^2}\right]
~\left[ 1 - \frac{(a \cosh^2 \alpha)^2}{r^2}\right]~dt^2
+
\frac{dr^2}
{\left[ 1 - \frac{(a \sinh^2 \alpha)^2}{r^2}\right]
~\left[ 1 - \frac{(a \cosh^2 \alpha)^2}{r^2}\right]} \nonumber
\\
&+& r^2 d\Omega_3^2~.
\label{ne5d}
\end{eqnarray}
The two horizons are at
\begin{equation}
r_+=a \cosh \alpha~~,  \;\;\;\;\;
r_-=a \sinh \alpha~~.
\end{equation}
It is evident that the extremal limit ($r_+\rightarrow r_-$) is achieved when the
boost parameter becomes indefinitely large, i.e. $\alpha \rightarrow
\infty$, and $a\rightarrow 0$ such that $ a e^\alpha $ is held fixed. 
Clearly, taking this limit does not change the Euclidean topology from
 $disc \times S^3$ to $annulus\times S^3$ as exact equality of the horizons 
is not achieved. The metric is of the proposed form 
(\ref{5d'}), coinciding with the metric (\ref{5d}) only in a limiting way.
Therefore the black hole metric considered in \cite{vafa} obeys the area 
law for entropy. The stability of the black hole is ensured by the fact that the Hawking temperature, 
$T_{H}=1/2\pi a\cosh^3 \alpha$, is infinitesimal.

>From the point of view of counting
microscopic degrees of freedom  using D-brane
techniques, it has been shown that the density of states varies 
continuously as a function of parameters \cite{near,malnon}.
In other words, there is no discontinuity at the BPS limit 
and the entropy calculated from the
degeneracy of BPS states is to be identified with $A/4G$ of
the corresponding black hole in the extremal limit. 

\section{Conclusions}

In this paper, we have shown that non-extremal and extremal black
holes are physically quite distinct objects and it is impossible
to transform the former to the latter by physical processes. 
These have different Euclidean topologies and hence 
they do not share the same entropy formula. Extremal black
holes have zero entropy as opposed to non-extremal black holes
which obey the area law. Since the degeneracy counting for
BPS saturated states follows the area law, 
we have proposed a form for the 
black hole metric which has the above property. 
We have justified this proposal by showing that this metric
corresponds to the extremal limit of the 
black hole solutions in string theory. 
Finally, we have pointed out that the BPS saturated 
D-brane configuration should actually be identified with
the corresponding black hole in the extremal limit (as opposed
to exactly extremal) such that the area law for entropy is valid. 
The interesting question remains as to what kind of
statistical interpretation can be given to the exactly extremal
black holes and whether they have a stringy interpretation. 
We hope to report on it elsewhere. 

\begin{center}
{\bf ACKNOWLEDGEMENTS}
\end{center}

We thank P. Majumdar for discussions and for carefully 
reading the manuscript 
and suggesting various improvements. We thank T. Jayaraman, 
T. Sarkar and G. Sengupta for discussions. We are grateful 
to A. Sen for useful correspondence. One of us (A.D.) would
like to thank F. Englert for discussions.

\end{document}